\documentclass[english,12pt]{article}
          \usepackage{babel}
          \usepackage[T1]{fontenc}
          \usepackage[latin2]{inputenc}
          \usepackage{pslatex}
\begin{document}

\title{\bf Multiperiodic RR~Lyrae Stars in
Omega Centauri}
    \author{P. Moskalik$^1$ ~and~ A. Olech$^1$}
\date{$^1$ Copernicus Astronomical Center, Bartycka 18, 00-716 Warszawa, Poland}
\maketitle

\begin{abstract}

We have conducted a systematic search for multiperiodic pulsations in
the RR~Lyrae-type stars of the galactic globular cluster $\omega$~Cen.
Secondary periodicities close to the primary pulsation frequency have
been detected in 17 out of 70 studied fundamental mode (RRab) pulsators
and in 35 out of 81 overtone (RRc) pulsators. Because of the observed
period ratios, these newly detected periodicities must correspond to
nonradial modes. Their beating with the primary radial pulsation leads
to a slow amplitude and phase modulation, commonly referred to as the
Blazhko effect. The incidence rate of Blazhko modulation in
${\omega}$~Cen RRab stars ($24\pm 5\%$) is similar to that observed in
the Galactic Bulge. In the case of $\omega$~Cen RRc stars, the incidence
rate of Blazhko effect is exceptionally high ($38\pm5\%$), more than 3
times higher than in any other studied population.

In addition to Blazhko variables, we have also identified two
RR~Lyr variables exhibiting first overtone/second overtone
double-mode pulsations, and a triple-mode High Amplitude
$\delta$ Scuti variable.

\end{abstract}

\section{Introduction}

Omega Centauri ($\omega$ Cen) is the largest globular cluster of the
Galaxy. It contains about one million stars of which almost 500 are known
to be variable (Kaluzny et al. 2004, Weldrake et al. 2007). Recently,
precise $B$ and $V$ CCD photometry of variables in $\omega$ Cen were
obtained by Cluster AgeS Experiment (CASE) team (Kaluzny et al. 2004).
Among others, they published precise light curves of 151 RR Lyr stars
belonging to the cluster. These light curves contained from 594 to 761
points and were collected in period from 1999 February 6/7 to 2000
August 9/10.

\section{Frequency analysis}

We conducted the search for periodicities in individual RR Lyr stars.
The analysis was performed in two ways.

We searched for additional signal in $\omega$~Cen RR~Lyrae stars
with the standard consecutive prewhitening technique. First, the
data were fitted with the single frequency Fourier sum of the form:

\begin{equation}
m(t) = A_{0} + \sum_{k} A_{k} \cos (2\pi k f_{0} t + \phi_{k})
\end{equation}

The pulsation frequency $f_{0} = 1/P_{0}$ was also optimized in the
fitting process. The fit Eq.(1) was then subtracted from the data
and the residuals were searched for secondary periodicities. This
was done with the Fourier power spectrum computed over the range of
$0-5$ c/d. If any additional signal was detected, a new Fourier
fit with all frequencies identified so far and with their linear
combinations was performed. All frequencies were optimized anew. The
new fit was subtracted from original time-series data and residuals
were searched for additional periodicities again. The process was
repeated until no new frequencies appeared. At this stage, we
performed data clipping, by rejecting all measurements deviating
from the fit by more than 5$\sigma$, where $\sigma$ is the
dispersion of residuals. After removing deviating datapoints, the
entire frequency analysis was repeated.

The second method, also based on the consecutive prewhitenings,
differed by treatment of the residua. We started from the ANOVA
periodogram (Schwarzenberg-Czerny 1996) of the raw light curve and
searched it for the peak frequency. Then the light curve was prewihtened
with the frequency just identified. Depending on the amplitudes we
removed up to 15 harmonics of the frequency from the the {\em
prewhitened data} of the previous stage. Next we recomputed the ANOVA
periodogram for the residuals and the whole procedure was repeated
recursively till no feature exceeded ANOVA value of 15.

It is worth to note that both methods produce, within errors, the same
results. The only discrepancies appeared when a real period and its alias
have similar height and one method preferred one peak and the second
the other one. In this case we have chosen the fit with smaller formal
error.

\section{Results}

We detected secondary periodicities close to the primary pulsation
frequency in 17 out of 70 fundamental mode pulsators (RRab or RR0
stars) and in 31 out of 81 first overtone pulsators (RRc or RR1
stars). The new frequencies are well-resolved within our dataset
and do not result from secular period variability. The derived
period ratios are incompatible with those of the radial modes.
Consequently, the newly detected secondary frequencies must
correspond to nonradial modes of oscillations. The beating between
these nonradial modes and the primary radial pulsation leads to a
slow modulation of the observed light curve, a phenomenon known as
the Blazhko effect.

\vspace{7.6cm}

\includegraphics{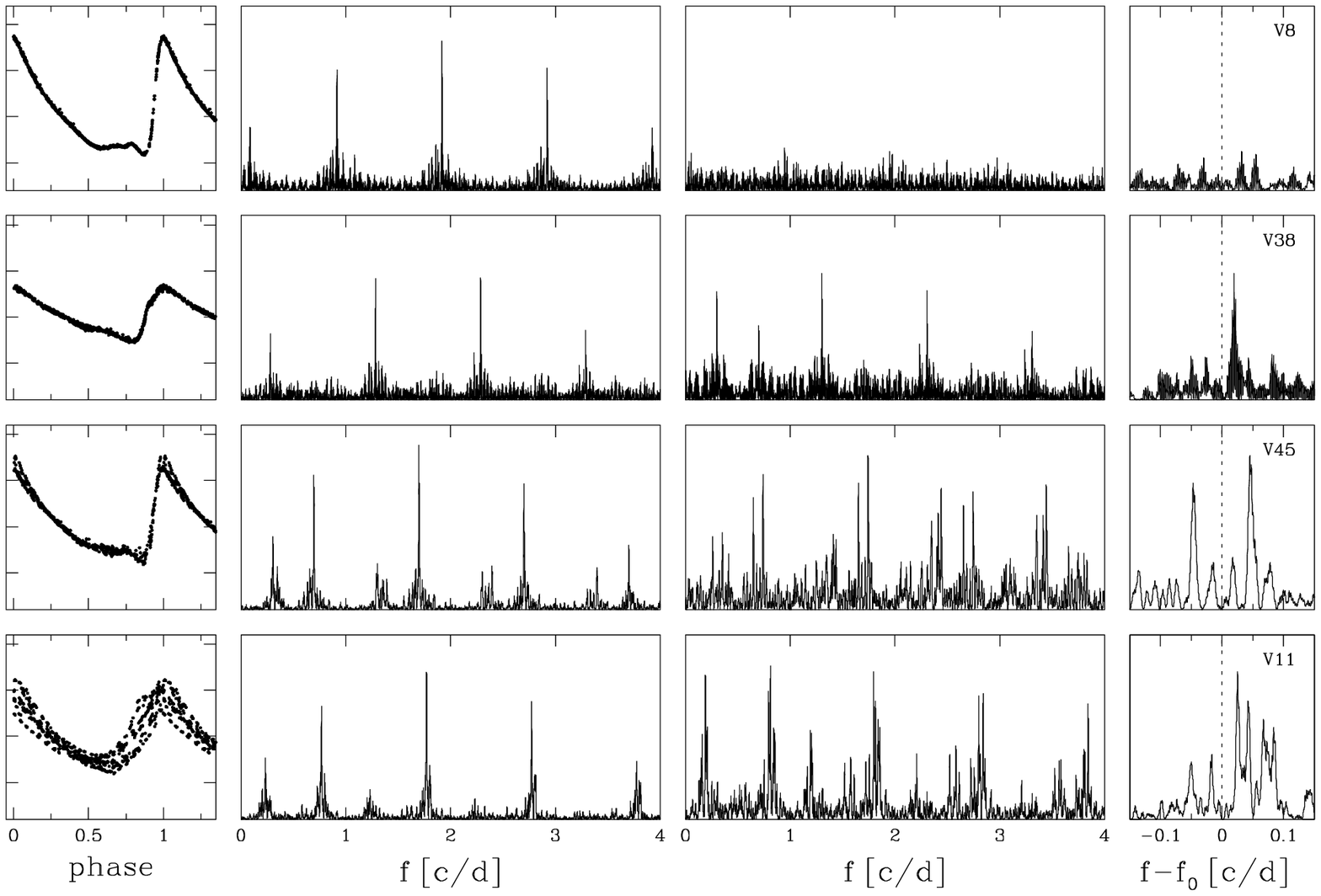}

   \begin{figure}[h]
      \caption{\sf 
Typical examples of different
frequency patterns encountered in the RRab variables. A true monoperiodic 
star is shown in the top panel of the plot.}
\end{figure}

In RRab variables two nonradial modes were detected in most cases.
Together with the primary peak, they form an equally spaced
triplet of frequencies, centered on the primary peak. Following
Alcock et~al. (2000) we call such stars RR0-BL variables. In 3
RRab stars only one nonradial mode is present (RR0-$\nu1$
variables). Finally, an even more complicated pattern was detected
in variable V11. In this object we found an equidistant triplet
plus an additional peak on the high frequency side of the primary
peak. In Fig.\thinspace 1 we display typical examples of different
frequency patterns encountered in the RRab variables. For
comparison, we show a true monoperiodic star in the top panel of
the plot.

In RRc variables usually only one nonradial mode was detected
(RR1-$\nu1$ stars). Equidistant triplets were found only in 4
variables (RR1-BL stars). However, more complicated patterns are
seen in RRc stars much more frequently than in the RRab stars. In
6 variables we detected {\it nonequidistant} triplets of modes. In
other 4 RRc stars, even richer frequency patterns with up to 7
nonradial modes were found.

Our findings are summarized in Tables\thinspace 1 and 2. In
$\omega$~Cen fundamental mode RR~Lyrae stars the incidence
rate of nonradial modes is $24.3\pm 5.1$\%. This is
twice the rate observed in the LMC, but identical to the rate in
the Galactic Bulge. For $\omega$~Cen overtone RR~Lyrae stars
the incidence rate of nonradial modes is $38.3\pm 5.4$\%, which is
exceptionally high. This is by far the highest rate among all
studied RRc populations, being 3 times higher than in the Bulge
and 4 times higher than in the LMC.

\bigskip

\noindent {\sc Table 1: Incidence rates of nonradial modes in RRab
stars in LMC, Galactic Bulge and $\omega$ Cen.}
\smallskip
\begin{center}
\begin{tabular}{|l|c|c|c|c|}
\hline
\hline
 & LMC & Bulge & Bulge & $\omega$ Cen \\
 & Alcock et al. & ~~~~Mizerski~~~~ & Collinge et al. & ~~~this work~~~ \\
 & (2003) & (2003) & (2006) & \\
\hline
\hline
RR0 & 6158 & 1942 & 1888 & 70 \\
\hline
RR0-$\nu 1$ & ~400 \hfill 6.5\%~ & ~243 \hfill 12.5\%~ & ~167 \hfill 8.8\%~ & ~~~3 \hfill 4.3\%~ \\
RR0-BL & ~331 \hfill 5.4\%~ & ~143 \hfill 7.4\%~ & ~282 \hfill 14.9\%~ & ~13 \hfill 18.6\%~ \\
RR0-oth & ~~~20 \hfill 0.3\%~ & ~~~86 \hfill 4.4\%~ & ~~~75 \hfill 4.0\%~ & ~~~1 \hfill 1.4\%~ \\
\hline
All NR & ~751 \hfill 12.2\%~ & ~472 \hfill 24.3\%~ & ~524 \hfill 27.8\%~ & ~17 \hfill 24.3\%~ \\
\hline
\hline
\end{tabular}
\end{center}
\bigskip\medskip

\noindent {\sc Table 2: Incidence rates of nonradial modes in RRc
stars in LMC, Galactic Bulge and $\omega$ Cen.}
\smallskip
\begin{center}
\begin{tabular}{|l|c|c|c|c|}
\hline
\hline
 & LMC & LMC & Bulge & $\omega$ Cen \\
 & Alcock et al. & Nagy \& Kov\'acs & ~~~Mizerski~~~ & ~~this work~~ \\
 & (2000) &  (2006)          & (2003) & \\
\hline
\hline
RR1 & 1143 & 1161 & 771 & 81 \\
\hline
RR1-$\nu 1$ & ~24 \hfill 2.1\%~ & ~~~46 \hfill 4.0\%~ & ~22 \hfill 2.9\%~ & ~17 \hfill 21.0\%~ \\
RR1-BL & ~28 \hfill 2.4\%~ & ~~~53 \hfill 4.6\%~ & ~30 \hfill 3.9\%~ & ~~~4 \hfill 4.9\%~ \\
RR1-oth & ~~~8 \hfill 0.7\%~ & ~~~13 \hfill 1.1\%~ & ~41 \hfill 5.3\%~ & ~10 \hfill 12.3\%~ \\
\hline
All NR & ~60 \hfill 5.2\%~ & ~112 \hfill 9.6\%~ & ~93 \hfill 12.1\%~ & ~31 \hfill 38.3\%~ \\
\hline
\hline
\end{tabular}
\end{center}
\bigskip

\section{Other interesting cases}

In the course of systematic frequency analysis of $\omega$~Cen
RR~Lyrae stars we found two double mode pulsators. These are the
first double mode RR~Lyrae variables identified in this globular
cluster. The period ratios in both stars are close to 0.80, which
implies pulsations in the first two radial overtones. Properties
of both variables are summarized in Table\thinspace 3.
\bigskip

\noindent {\sc Table 3: FO/SO double mode RR~Lyrae stars in $\omega$~Cen}
\smallskip
\begin{center}
\begin{tabular}{|l|c|c|c|c|c|}
\hline
\hline
Star &  $P_1$ [day] &  $P_2$ [day] &  $P_2/P_1$ & $A1$ [mag] & $A2$ [mag]\\
\hline
V10   & ~0.3749761~ & ~0.2991768~ & ~0.797856~ & ~0.18305~ & ~0.00668~ \\
V145  & ~0.3732140~ & ~0.2986517~ & ~0.800216~ & ~0.20481~ & ~0.00894~ \\
\hline
\hline
\end{tabular}
\end{center}
\bigskip

RR~Lyrae stars pulsating in the first and second overtone (FO/SO
variables) have been previously discovered in the LMC (Alcock
et~al. 2000). However, Soszy\'nski et~al. (2003) have noticed,
that all these stars are about 1\thinspace mag brighter than
typical RR~Lyrae stars in the LMC and suggested that they might
belong to short-period double mode Cepheids. In contrast, the two
FO/SO double mode pulsators discovered in $\omega$~Cen belong to
the RR~Lyrae population of the cluster without any doubt.

In one object, V168, we discovered triple mode pulsations. This
star varies with $P_0=0.3212979$\thinspace day,
$P_1=0.2465336$\thinspace day and $P_2=0.1967626$\thinspace day.
The resulting period ratios are $P_1/P_0=0.767305$ and
$P_2/P_1=0.798117$. These ratios indicate that V168 pulsates in
the fundamental mode and the first two radial overtones (FU/FO/SO
pulsator). The value of $P_1/P_0$ is too large for RR~Lyrae stars,
but it is rather typical for High Amplitude $\delta$~Scuti stars
(HADS), especially for those with longest periods (Poretti et~al.
2005). We note in passing, that period ratios of V168 are very
similar to ratios observed in two other triple mode HADS, V829~Aql
(Handler et~al. 1998) and V823~Cas (Jurcsik et~al. 2006).

\bigskip
\section*{References}

\noindent Alcock C., Allsman R., Alves D.R., et al., 2000, {\sl ApJ}, 
{\bf 542}, 257

\noindent Alcock C., Alves D.R., Becker A.,  et al., 2003, {\sl ApJ}, 
{\bf 598}, 597

\noindent Collinge M., Sumi, T., Fabrycky D., 2006, {\sl ApJ}, {\bf 651}, 197

\noindent Handler G., Pikall H., Diethelm R., 1998, IBVS 4549

\noindent Jurcsik J., Szeidl B., V\'aradi M., et~al., 2006, {\sl Astron.
Astrophys.}, {\bf 445}, 617.

\noindent Kaluzny J., Olech A., Thompson I.B., et al.,  2004, {\sl Astron. 
Astrophys.}, {\bf 424}, 1101

\noindent Mizerski T., 2003, {\sl Acta Astron.}, {\bf 53}, 307

\noindent Nagy A., Kov\'acs G., 2006, {\sl Astron. Astrophys.}, {\bf 454}, 257

\noindent Poretti E., Su\'arez J.C., Niarchos P.G., et~al., 2005,
{\sl Astron. Astrophys.}, {\bf 440}, 1097.

\noindent Schwarzenberg-Czerny A., 1996, {\sl ApJ Letters}, {\bf 460}, L107

\noindent Soszynski I., Udalski A., Szymanski M. et al., 2003, {\sl Acta Astron.},
{\bf 53}, 93

\noindent Weldrake D.T.F., Sackett P.D., Bridges T.J., 2007, {\sl Astron. J},
{\bf 133}, 1447

\vspace{0.5cm}
\begin{flushleft}
{\bf Acknowledgments.} ~This work was supported by MNiI grant
number 1 P03D 011 30 to P.M.
\end{flushleft}

\end{document}